\documentclass[12pt]{article}
\title{New (virtual) Physics in the Era of the LHC}
\author{M.I. Vysotsky \\
ITEP, Moscow, Russia}
\date{}
\begin{document}
\maketitle

\begin{abstract}
A simple extension of the Standard Model demonstrates that
New Physics
non-reachable through direct production at LHC can
induce up to 10\% corrections to the length of the unitarity
triangle side, extracted from $\Delta m_{B_d}$.
\end{abstract}

Let us imagine the worst scenario: the only new particle found at
the LHC will be the Higgs boson of the Standard Model (SM). A
natural question arises: is it possible to find traces of New
Physics in low energy observables without observing the production
of new particles at LHC? Another facet of this question: What
changes of the unitarity triangle can be produced by such
particles? This is the problem we will focus on.

In order to
influence the quark weak currents the new particles should be
strongly interacting. The natural example would be the fourth
quark-lepton family: the fourth generation quarks deform
unitarity triangle into unitarity quadrangle.
However since the sequential fourth
generation gets masses through Higgs mechanism, its quarks cannot
be heavier than 1 TeV: so, they will be directly produced at LHC.
That is why
the heavy particles we are looking for should get their masses
from a different source. So their contributions to low energy 
observables decouple, being suppressed as $(\eta /M)^2$, where
 $\eta =
246$ GeV is the Higgs boson neutral component expectation value
and $M$ characterizes new particles masses, $M \ge 5$ TeV in
order to avoid their production at LHC. These 1\% corrections 
are too small to be detected taken into acount relatively low 
accuracy of theoretical formulas. Nevertheles we manage to find
a model where corrections are enhanced and can be detected.

Let us study the extension of SM by
$SU(2)_L$ singlet heavy Dirac quark $Q$ with electric charge
$+2/3$ which mixes with the top
quark. Recently
the constraints from the $B \to X_s \gamma$ branching ratio 
and electroweak precision observables in this model
have been studied \cite{A}.

 The model is described by the following lagrangian:
\begin{equation}
{\cal L} = {\cal L}_{SM} - M \bar Q^\prime Q^\prime + \left[\mu_R
\bar Q^\prime_L t^\prime_R + \frac{\mu_L}{\eta/\sqrt 2} H^+ \bar
Q_R^\prime\left(\begin{array}{l} t^\prime \\ b^\prime
\end{array}\right)_L + c.c. \right] \;\; ,
\label{1}
\end{equation}
where ${\cal L}_{SM}$ is the SM lagrangian, $M$, $\mu_R$ and
$\mu_L$ are the parameters with the dimension of mass.
The term proportional to $M$ contains Dirac mass of the field
$Q^\prime$ which is primed since it is not a state with a definite
mass due to mixing with $t$-quark. The term proportional to
$\mu_R$ describes the mixing of two $SU(2)_L$ singlets: $Q_L^\prime$
and $t_R^\prime$, the latter being the right component of
$t$-quark field in the Standard Model (in the absence of terms in square brackets).
Finally, the term proportional to $\mu_L$ describes mixing of a
weak isodoublet with $Q^\prime$. An upper component of this
isodoublet is the left component of the field $t^\prime$ which
would be $t$-quark without the terms in square brackets:
\begin{equation}
t_L^\prime = U_{t^\prime t^{\prime\prime}}^L t_L^{\prime\prime} +
U_{t^\prime c^\prime}^L c_L^\prime + U_{t^\prime u^\prime}^L
u_L^\prime \;\; , \label{2}
\end{equation}
where $t^{\prime\prime}$, $c^\prime$ and $u^\prime$ are the
primary fields of SM lagrangian, while $U_{ik}^L$ are the matrix
elements of matrix $U^L$ which transforms the primary fields
$c_L^\prime$ and $u_L^\prime$ to the left-handed components of the
mass eigenstates $c$ and $u$ and field $t_L^{\prime\prime}$ to the
field $t_L^\prime$ which would be the left-handed component of the
top quark in the case $\mu_L = \mu_R = 0$.  We do not mix $Q-$quark with
$u-$ and $c-$ quarks in order to avoid FCNC which may induce too
large $D^0 - \bar D^0$ oscillations.

One can easily see that
the lower component of the isodoublet is 
the
combination of the down quark fields with definite masses
rotated by CKM matrix $V$:

\begin{equation}
b_L^\prime = V_{tb} b_L + V_{ts} s_L + V_{td} d_L \;\; . \label{3}
\end{equation}

In order to find the states with definite masses which result
from $t^\prime$--$Q^\prime$ mixing, the following matrix should be
diagonalized:
\begin{equation}
(\overline{t_L^\prime} \overline{t_R^\prime} \overline{Q_L^\prime}
\overline{Q_R^\prime}) \left( \begin{array}{cccc} 0 & m_t & 0 &
\mu_L \\ m_t & 0 & \mu_R & 0 \\ 0 & \mu_R & 0 & -M \\ \mu_L & 0 &
-M & 0 \end{array} \right) \left( \begin{array}{l} t_L^\prime \\
t_R^\prime \\ Q_L^\prime \\ Q_R^\prime \end{array} \right) \;\; ,
\label{4}
\end{equation}
where $m_t$ is the mass of $t$-quark in SM. For the squares of
masses of the eigenstates  we get:
\begin{eqnarray}
2(\lambda^2)_{t,Q} & = & M^2 + \mu_R^2 + \mu_L^2 + m_t^2  \mp \\ &
\mp & \sqrt{(M^2 + \mu_R^2 + \mu_L^2 + m_t^2)^2 - 4M^2 m_t^2 -
4\mu_L^2 \mu_R^2 + 8m_t \mu_R \mu_L M} \;\; , \nonumber \label{5}
\end{eqnarray}
and the eigenstates look like (in what follows we put $m_t =0$\footnote{We did it in order to
simplify the formulas a bit; however this can be suggested as an
explanation of heaviness of top: $t$-quark massless in SM gets all
its mass due to mixing with heavy $Q$.}):
\begin{equation}
t = t_L^\prime + (1-\frac{\lambda_t^2}{\mu_L^2}) \frac{\mu_R
\mu_L}{\lambda_t M} t_R^\prime + \frac{\mu_L}{M} (1-
\frac{\lambda_t^2}{\mu_L^2}) Q_L^\prime + \frac{\lambda_t}{\mu_L} Q_R
^\prime
\;\; , \label{6}
\end{equation}
\begin{equation}
\lambda_t = \frac{\mu_R \mu_L}{M}\left(1-\frac{\mu_R^2 +
\mu_L^2}{2M^2}\right) + O(\frac{1}{M^5}) \;\; , \label{7}
\end{equation}
\begin{equation}
Q = Q_R^\prime + (-\frac{\lambda_Q}{M} + \frac{\mu_L^2}{\lambda_Q M}) Q_L
^\prime
+ \frac{\mu_L}{\lambda_Q} t_L^\prime+ \frac{\mu_R}{M}
\left(\frac{\mu_L^2}{\lambda_Q^2} -1 \right) t_R^\prime \;\; , \label{8}
\end{equation}
\begin{equation}
\lambda_Q = -M + O\left(\frac{1}{M}\right) \;\; , \label{9}
\end{equation}
the normalization factors of the quark fields which
should be taken into account when calculating Feynman diagrams
are omitted.

Now we are ready to discuss the flavor changing quark
transitions.\\ $\bar t_R (b_L, d_L, s_L)H^+$ transition vertex
originates in our model from $Q_R$ admixture in the $t$-quark wave
function:
\begin{eqnarray}
\frac{\mu_L}{\eta/\sqrt 2}
\frac{\lambda_t/\mu_L}{\sqrt{\frac{\mu_L^2 \mu_R^2}{\lambda_t^2
M^2}\left(1- \frac{\lambda_t^2}{\mu_L^2}\right)^2 +
\frac{\lambda_t^2}{\mu_L^2}}} \bar t_R b_L^\prime H^+ = \nonumber
\\
= \frac{\lambda_t}{\eta/\sqrt 2}
\frac{1}{\sqrt{1+(\frac{\mu_L}{M})^2
\left(1-\frac{\lambda_t^2}{\mu_L^2}\right)^2}} \bar t_R b_L^\prime
H^+ \;\; , \label{10}
\end{eqnarray}
that is why up to the corrections $\sim(\mu_L/M)^2$ the box
diagrams for $B_{d,s} - \bar B_{d,s}$, $K^0 - \bar K^0$
transitions with the intermediate $t$-quarks are the same as in
SM\footnote{Since $H^+$ is the longitudinal $W^+$-boson
polarization its interaction is the same as that of $W^+$ and the
square root in the denominator from $(t_R, Q_R)$ proper
normalization equals that for $(t_L, Q_L)$ component.}.

How large can the term $(\mu_L/M)^2$ be? According to Eq.(\ref{1})
$\mu_L$ cannot be larger than 500 GeV: in the opposite case we
will be out of the perturbation theory domain and no calculations
can be trusted. That is why trying to have the largest possible
deviations from SM we will take $\mu_L = 500$ GeV in what follows.
The smallest value of $M$ which will prevent the production of
$Q$-quarks at LHC is about 5 TeV, and we will use it in order to
maximize deviations  from SM (consequently $\mu_R = m_t M/\mu_L
\approx 1.7$ TeV). At one loop level Q-quark contributes to 
$Z \to b \bar b$ decay. The analysis of the experimental data
made in \cite {A}  lead to $\mu_L/M \le 0.4$, and we are on
the safe side. The constraint from $B \to X_s \gamma$ decay is
even weaker.
The box with two
intermediate $t$-quarks is equal to that in SM with
$(\mu_L/M)^2\approx 1$\% accuracy. Theoretical uncertainties in
matrix elements calculations do not allow to detect 1\% deviation
from SM results.

Our model generates extra contributions to $\Delta F = 2$
four-fermion operators due to the boxes with intermediate
$Q$-quarks. The boxes with $H^+$ exchanges generate leading 
contributions in the
limit $m_t, M \gg M_W$. The box with one $t$-quark
substituted by $Q$ gives coefficient $\sim G_F^2 m_t^2 (\mu_L/M)^2
\ln (M/m_t)^2$: once more the correction is damped by the factor
$(\mu_L/M)^2 \approx 1\%$ relative to the SM contribution.

The largest correction comes from the box with two intermediate
$Q$-quarks:
\begin{equation}
\left(\frac{|\mu_L|}{\eta/\sqrt 2}\right)^4 \frac{1}{M^2} (\bar b_L
\gamma_\mu d_L) (\bar b_L \gamma_\mu d_L) \;\; , \label{11}
\end{equation}
where as an example we present the operator responsible  for $B_d
- \bar B_d$ oscillations. In this way we get:
\begin{equation}
\frac{{\rm box} (QQ)}{{\rm box} (tt)} \approx \frac{\mu_L^4}{m_t^2
M^2} \approx 10\% \;\; . \label{12}
\end{equation}

The explicit formula which takes into account $(tt)$ and $(QQ)$ boxes
can be easily obtained from that of SM \cite{VIL}:
\begin{eqnarray}
\Delta m_{B_d}& = & \frac{G_F^2 B_{B_d} f_{B_d}^2}{6\pi^2}
m_B\left[m_t^2 I\left(\frac{m_t^2}{M_W^2}\right)+ M^2
\left(\frac{|\mu_L|}{M}\right)^4
I\left(\frac{M^2}{M_W^2}\right)\right] \eta_B |V_{td}|^2 \; ,
\nonumber \\ I(\xi)& = & \left\{\frac{\xi^2 - 11\xi +4}{4(\xi
-1)^2} - \frac{3\xi^2 \ln\xi}{2(1-\xi)^3}\right\} \;\; ,
\nonumber
\\ I(0) & =  & 1 \; ; \;\; I\left(\frac{m_t^2}{M_W^2}\right)
\approx 0.55 \; ; \;\; I(\infty) = 0.25 \;\; . \label{13}
\end{eqnarray}

In conclusion we have found a simple extension of SM with one
additional heavy quark $Q$, $M_Q \approx 5$ TeV (non-reachable by
direct production at LHC), in which the corrections to CP
violating factor $\varepsilon$ in $K - \bar K$ transitions and the
values of $\Delta m_{B_d}$ and $\Delta m_{B_s}$ are universal and
can reach 10\%. We demonstrate that even with no new particles
found at LHC one cannot claim that the Unitarity Triangle is
universal and unambiguously extractable from different observables
with the accuracy better than 10\%. In our case the triangle 
determined by angles found from CP-asymmetries in B-decays and by one
side $(V_{cb}^* V_{cd})$ has the value of  side $(V_{tb}^* V_{td})$
which, being substituted into the SM expression for $\Delta
m_{B_d}$, produces the number smaller than the one extracted from
the measurement of the $B_d - \bar B_d$ oscillation frequency by
$\approx$10\%. However, to detect this discrepancy one needs to
have an accuracy in the value of the product $f_{B_d}^2 B_{B_d}$
better than 10\% (the present day accuracy is about 2 times worse
\cite{3}).

In recent paper \cite{BB} the contribution to $\Delta m _{B_{d,s}}$
due to singlet heavy fermion with electric charge $+2/3$ has been
studied. The analized model is motivated by a Little Higgs scenario.
In this scenario our factor $\mu_L$ is substituted by $x_L \eta$,
where $0 \leq x_L \leq 1$ \cite{Logan}. That is why even for $x_L=1$
correction to $\Delta m _{B_{d,s}}$ is damped by the factor $2^4=16$
compared to our value.

This paper has arrived as the answer to Andrei Golutvin's question; I
am grateful to A. Golutvin and V. Novikov for valuable comments.
This work was partly supported by grants RFBR 05-02-17203 and
HSh-2603.2006.2.

\end{document}